\documentclass[superscriptaddress,aps,twocolumn,prl,showkeys,showpacs,author-numerical]{revtex4-1}
\usepackage{graphicx}
\usepackage{dcolumn}
\usepackage{bm}
\usepackage{hyperref}
\usepackage{color}
\usepackage[mathlines]{lineno}

\begin{document}


\title{Slow Relaxation and Aging Phenomena at the Nanoscale in Granular Materials} 
\author{V.\,Y. Zaitsev}
\email{vyuzai@hydro.appl.sci-nnov.ru}
\affiliation{ 
Institute of Applied Physics, RAS, Uljanova St. 46, 603950, Nizhny Novgorod, Russia}
\affiliation{%
LUNAM Universit\'e, Universit\'e du Maine, CNRS, LAUM UMR 6613, av.~O.~Messiaen, 72085 Le Mans, France
}%
\affiliation{%
Nizhny Novgorod State University, avenue~Gagarina 23, 603950, Nizhny Novgorod, Russia
}
\author{ V.\,E. Gusev} \author{V. Tournat}
\affiliation{%
LUNAM Universit\'e, Universit\'e du Maine, CNRS, LAUM UMR 6613, av.~O.~Messiaen, 72085 Le Mans, France
}%

\author{P. Richard}
\affiliation{%
LUNAM Universit\'e, IFSTTAR, site\,de\,Nantes, Route\,de\,Bouaye\, CS4, 44344 Bouguenais Cedex, France
}%
\date{\today}
\begin{abstract}
Granular matter exhibits a rich variety of dynamic behaviors, for which the role of thermal fluctuations is usually ignored. Here we show that thermal fluctuations can pronouncedly affect contacting nanoscale asperities at grain interfaces and brightly manifest themselves through the influence on nonlinear-acoustic effects. The proposed mechanism based on intrinsic bistability of nanoscale contacts comprises a wealth of slow-dynamics regimes including slow relaxations and aging as universal properties of a wide class of systems with metastable states.
\end{abstract}


\maketitle

%
%
\textit{Introduction.--} 
Slow relaxation phenomena in granular systems \textcolor{black}{are of considerable
interest to understand the physics of complex glassy-type systems \cite{Amir2011}, for
which they act as macroscopic analogs of ensembles of atoms and molecules}~\cite{Richard2005}. Granular systems are also crucial in many industrial and
geophysical applications.
Particularly, sufficiently ample understanding of nonlinear dynamics of individual contacts is required to interpret such intriguing and poorly understood phenomena as triggering of earthquakes by elastic waves with amplitudes significantly smaller than the damage threshold for rocks \cite{PAJ-Nat2005, PAJ-Nat2008}.
Compared to the widely \textcolor{black}{studied  slow macroscopic ({i.e.}, grain-scale) rearrangements causing compaction of granular materials~\cite{Knight1995,Richard2003,Richard2005,Ribiere2005b,Inserra2008},} processes 
at nanoscale potentially driven by weak strains, including natural thermal fluctuations, are much less studied. 
The latter are reasonably believed irrelevant to 
grain rearrangements during
compaction and jamming-unjamming transitions \cite{Richard2005,Deboeuf2005}. 
\textcolor{black}{However,} using appropriate acoustic techniques, spontaneous thermally activated 
\textcolor{black}{nanoscale} processes 
can also be 
macroscopically observed.
In particular, observations of slow relaxation of the elastic modulus 
in laboratory samples with cemented granular structure \cite{Tencate2000} as well as similar effects in field measurements in sandy soil on a scale of $\sim10$\,m \cite{averb-leb2009} are known. High-intensity acoustic ``conditioning'' \cite{Tencate2000}  or a mechanical impact \cite{averb-leb2009} ruptured 
the weakest bonds and produced perturbations in the 
elastic moduli of order $10^{-6}-10^{-3}$ that were rather problematic to monitor. 
%
To overcome this experimental difficulty, \textcolor{black}{a parameter dominated by the
weak-bond-network rather than the stable material skeleton is highly
desirable.}\\
\textcolor{black}{Here, we report (i) implementation of such an unconventional experimental approach,
(ii) a model of individual-contact bistability having essentially new features compared with conventional ones discussed for AFM tips and adhesion hysteresis, and (iii) }
\textcolor{black}{results of numerical simulations of collective behavior of such bistable contacts. These results capture essential observed features, in particular,} 
\textcolor{black}{
the abrupt breaking  of the nanoscale contacts, their slow  post-shock restoration and the peculiar aging of the system, and the damage accumulation produced by repeated weak perturbations.}


\textit{Methods.--}
 Experimentally, we use an acoustic  (usually $P$-wave) component  produced by the own material nonlinearity, which is strongly dominated by the contributions of the weakest-contact fraction \cite{Zaitsev1995, Tournat2004}.  Thus,  amplitude variations of the nonlinear component, characterize temporal variations in the amount of  contributing weak contacts. Compared with intact homogeneous solids, nonlinearity of granular packings is giant and can be observed much easier. Feasibility of such 
{nonlinear-acoustic sounding} 
was demonstrated
in Ref.~\cite{Zaitsev2005} using the 
nonlinear cross-modulation technique to
monitor structural perturbations in granular material bulk induced by weak mechanical shocks. Another, practically simpler, nonlinear-demodulation technique was successfully applied for  studying fine structural changes --avalanche precursors-- in slowly tilted granular packings \cite{Zaitsev2008, JSTAT2010}. 

Here, 
the sounding technique \cite{Zaitsev2008, JSTAT2010}, combined with pulse-type perturbations \cite{Zaitsev2005},
is used to study slow  relaxation of the weak-bond network in granular material  
with particular attention to   aging of material 
undergoing repeated perturbations \cite{JSTAT2010}.      
\textcolor{black}{We use random packings of glass beads 1 and 2 mm  in diameter placed in a container $5-10$~l in volume}, to which a small electromagnetic shaker is attached.
It produces perturbing pulses that \textcolor{black}{are much weaker compared to typical conditions of tap-induced compaction \cite{Knight1995,Richard2003,Richard2005,Ribiere2005b} and surely do not cause macroscopic grain rearrangements}. The strain amplitude of the pulses varied in different measurements from about $10^{-7}$ to $10^{-6}$ and their duration is  20 ms. The primary amplitude modulated wave  is at strain amplitude $\tilde{\varepsilon}_A\sim 10^{-8}-10^{-7}$ (see also Ref.~\cite{Suppl-1}). Unlike observations of the primary
wave \cite{Tencate2000,averb-leb2009} dominated by the medium skeleton, we use the demodulated component that is strongly dominated by the weakest contacts in the material. This is confirmed by the fact that such moderate shocks with nanometer and even subnanometer displacements, which are able to break only the weakest contacts, can cause several-times drops in the demodulated-component  amplitude.
Thus, such drops  are proportional to the number of shock-ruptured weak contacts (see also Ref.~\cite{Zaitsev2008}), and the slow relaxation of the nonlinear-component amplitude reflects how those contacts are restored.

\begin{figure}

\includegraphics[width=8.5 cm]{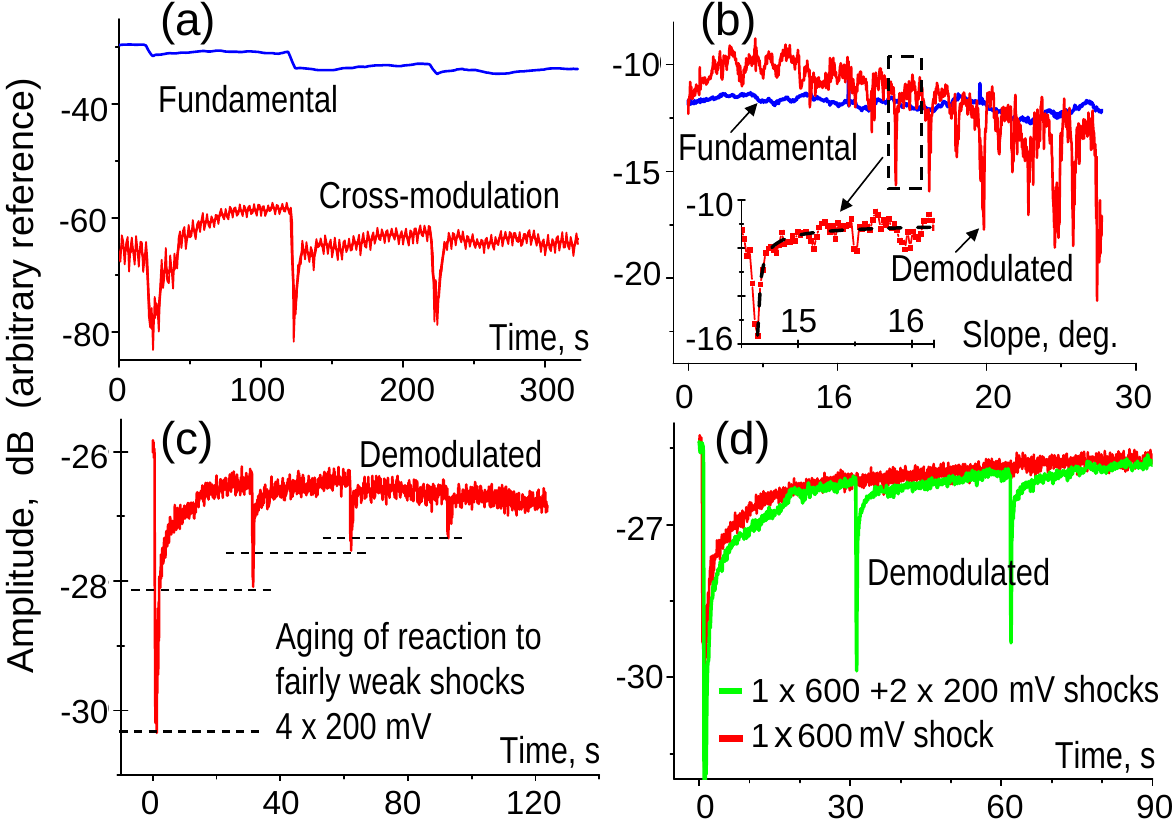}

\caption{(color online). 
Monitoring of slow relaxations via nonlinear-acoustic signal: (a)  after perturbing shocks using cross-modulation technique \cite{Zaitsev2005}; (b)  due to internal microslips in slowly tilted granular packings observed using demodulation  technique \cite{Zaitsev2008,JSTAT2010};  (c)  similarly observed ``aging'' of the system response to a series of identical fairly weak 
shocks, and  (d)  relaxation after one stronger shock and two weaker 
 shocks that perturb barriers with essentially different energies. \textcolor{black}{Notice much weaker perturbation of fundamental (linear) components shown in panels (a) and (b) for comparison.} }
\label{fig-raw-examp}
\end{figure}

Figure\,\ref{fig-raw-examp} shows examples of slow relaxation in granular materials, observed \textcolor{black}{via the amplitude of their nonlinear-acoustic response.} \textcolor{black}{If time $t$ is counted from the shock endings, the latter amplitude demonstrates} power-law rates close to $1/t^{1+n}$ with $|n|\ll 1$, i.e., close to log-time behavior corresponding to $n=0$. Plots (a) and (c) demonstrate peculiar weakening of the material reaction to series of identical taps, i.e., a kind of ``aging.'' 
\textcolor{black}{Besides, plots (a) and (b) show that the nonlinearity-produced signal is much more sensitive to the state of weakest contacts than the 
fundamental component variability.}
\textcolor{black} {
 Concerning the gradual relaxation of the shock-induced perturbations, we note that} even if the probing signal is switched off just after the shock and switched on after a pause, the nonlinearity of the material restores spontaneously.
This shows that the influence of the  probing 
wave  does not dominate the effect, 
although high-intensity acoustic strains (say $10^{-5}$) may 
perturb the weak bonds~\cite{Tencate2000,Suppl-1,Tournat2009}.

 \begin{figure}
\includegraphics[width=8.9 cm]{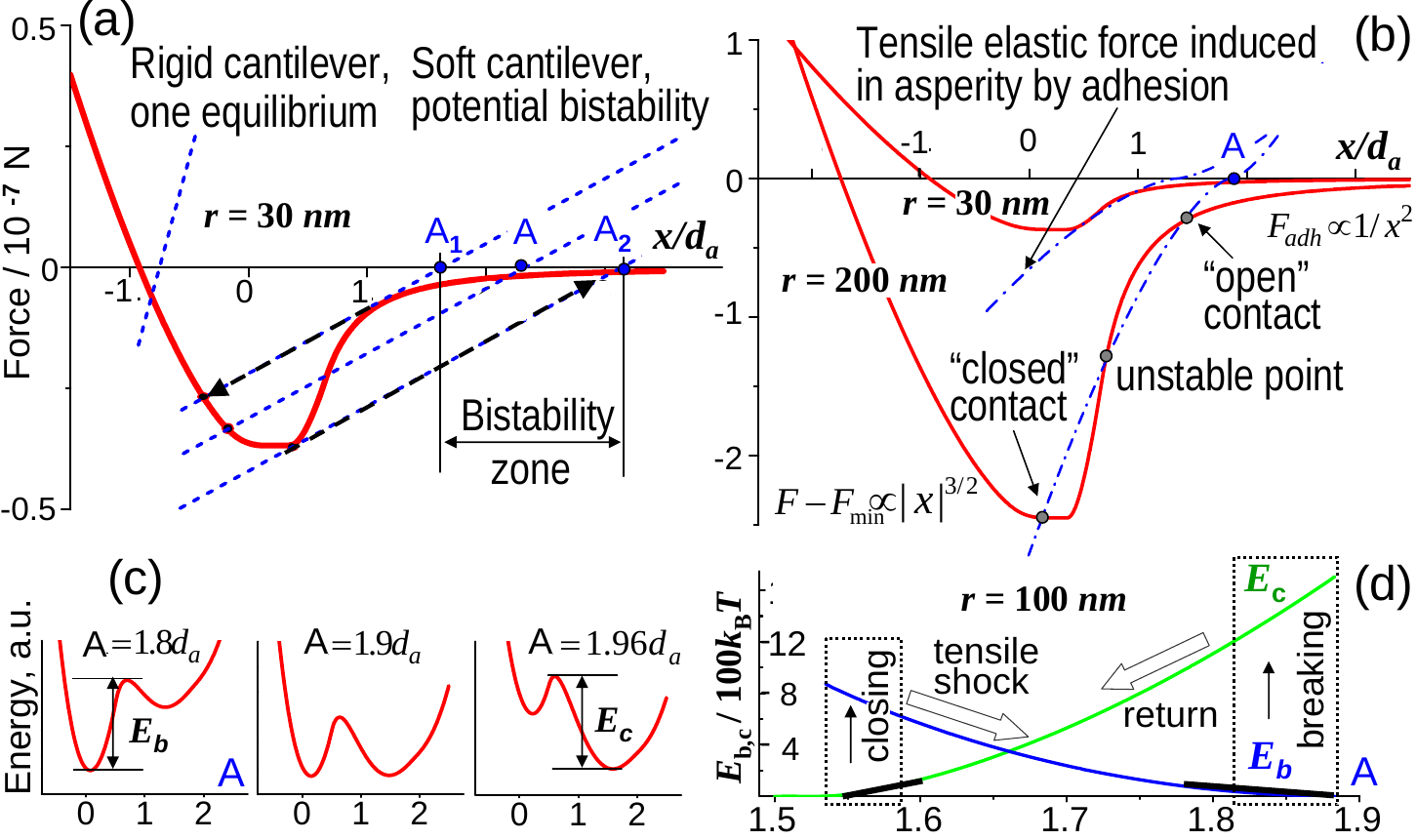}

\caption{(color online). \textcolor{black}{Schematic elucidation of the contact-bistability origin.} (a) Adhesion or compression force (solid line) for a small asperity ($r=30$\,nm)  and equilibrating elastic force of a cantilever (dashed lines) which jointly are able to produce bistable equilibria \textcolor{black} {if the cantilever is soft enough (the most-right and most-left intersection points).}
The elastic parameters and surface energy used are those of glass. (b) Elucidation that Hertzian-like elastic {\it tensile} force for sufficiently large contacts subjected to short-range adhesion forces from an opposite solid surface can also form bistable equilibria. (c) Corresponding two-minimum potential wells, for which the energy barriers 
 $E_{b,c}$ exhibit opposite trends as a function of separation $A$ normalized to atomic size $d_a$. (d) {Energy barriers $E_b(A)$ and $E_c(A)$ that
should be overcome to break ``closed" contacts and close ``open'' ones. Dashed rectangles 
show low-energy regions in which thermal fluctuations can induce jumps to the opposite equilibrium. }}
\end{figure}

{\textit{Mechanism.}--} Under room temperature $T\sim300$\,K the characteristic thermal energy  $k_{B}T $ 
($k_B$ being the Boltzmann constant) unambiguously indicates that thermal fluctuations cannot affect the state of visible, even weakly loaded, macroscopic contacts \textcolor{black}{usually considered   in granular matter modeling \cite{Deboeuf2005,Staron2002}.}  Thus only 
nanoscale surface asperities 
(from tens to hundreds of nanometers) 
can be considered as candidates of bistable structural elements potentially sensitive to thermal fluctuations. To understand the origin of their bistability,  the analogy with the bistable behavior of tips in atomic-force microscopy (AFM) is very useful.  For a tip already compressed by the contacting solid, the 
Hertzian force is repulsive, whereas the tip yet approaching a solid surface experiences the influence of short-range attraction forces.
This attraction force for an AFM tip approaching another solid is equilibrated by the elasticity of the cantilever [dashed lines in Fig.\,2(a)].  If the cantilever is soft enough, a bistability zone \textcolor{black} {for the initial position $A_1\le A \le A_2$ of the unstressed cantilever} can appear~\cite{Bhushan} as illustrated in Fig.\,2(a). In this zone, 
the cantilever can equilibrate the attraction force at two positions of the tip, ``closed'' and ``open.''
Note that,  in the latter \textcolor{black}{position}, the attraction force is almost absent. If the cantilever is moved forth and back, peculiar hysteretic jumps between the two positions occur [arrows in Fig.\,2(a)].
Inside the bistability zone 
sufficiently strong thermal fluctuations can cause transitions between the two equilibrium states.

At first glance, for 
nanoscale asperities at grain surfaces, there is no ``soft cantilever'' to create bistable equilibria like in AFM. 
\textcolor{black}{However, as argued in Ref.~\cite{ContMech}, the elastic energy stored in compressed
contacts scales like $h^{5/2}$, $h$ being the displacement of the contact
apex. Accordingly, the elastic force $F_{\rm{comp}}$ follows the Hertzian law
$F_{\rm{comp}} \propto h^{3/2}$.}
But the same arguments applied to a contact apex displaced by the value $|h|=|x-A|$ due  to a localized attractive 
force also lead to the appearance of an elastic force  $F_{\rm{tens}}\propto| x-A|^{3/2}$ that equilibrates the attraction [dash-dotted curves in Fig.\,2(b)].
Unlike AFM cantilever elasticity, this elastic force is nonlinear; i.e., initially it can be sufficiently soft  to create the second (distant) equilibrium position  for the contact tip. For AFM tips with the typical radius $r\alt10$\,nm, however, this ``nonlinear spring'' hidden inside the tip is insufficiently soft  relative to adhesion, so that only an in-sequence connected soft cantilever can create the second potential minimum. But for a larger contact radius $r\agt30-50$\,nm, due to a different dependence on $r$ for the attraction and the ``hidden spring'', the latter  becomes sufficiently soft  relative to the adhesion force [compare the curves for  $r=30$\,nm and $r=200$\,nm 
in Fig.\,2(b)].
Thus, \textcolor{black}{for the larger contacts,} bistable equilibria 
 can appear  in a finite range of separations between the asperity \textcolor{black}{apex} and the opposite surface without \textcolor{black}{the necessity of an artificial soft spring or cantilever}. Figure\,2(c) schematically shows how the resulting two-minima potential evolves with the initial separation $A$. These representations suggest physically clear interpretation to the well-known \cite{deryagin1987surface} transition from the so-called Derjaguin-Muller-Toporov (DMT) model of very small contacts \textcolor{black}{not having bistability}
 to the Johnson-Kendall-Roberts (JKR) model for larger contacts exhibiting adhesion hysteresis~\cite{deryagin1987surface}.  The latter can be viewed as a special case of mechanical hysteresis, like for AFM tips, but \textcolor{black}{arising without an artificial soft cantilever.}   
%

\textcolor{black}{For elastic and surface energy typical of glasslike materials,
contacts with  $r\sim10^2$\,nm and $T \sim 300$K exhibit narrow regions near
the boundaries of the bistability zone, where one of the potential wells [$E_b$ or $E_c$, see Fig.\,2(b)]
 is $10^2$ to $10^4$ times larger than $k_B T$, while
the other is \textcolor{black}{of order} $10^1 k_B T$.} 
Thus, near the left boundary of the bistability region the closed state is much stabler, whereas the open one is metastable. Near the right boundary, the situation is the opposite.  The metastable equilibrium 
energy  is comparable with that of thermal fluctuations. So, they are able to induce jumps to the opposite stabler state with characteristic waiting times $\tau_0 \exp(E_{b,c}/k_{B} T)$ according to the Arrhenius law, where the attempt time  for nanometer-scale tips of the asperities can reasonably be 
$\tau_0\sim10^{-12}$s. 


Direct AFM inspection of the glass-bead surfaces confirmed the presence of numerous asperities about $10^2$nm in radius and $20-50$nm in height (see Ref.~\cite{Suppl-2}) consistent with the values reported in~\cite{Divoux2008}. A single macrocontact between two grains leads to  $10^3-10^4$ microasperities.
Even if 1\% of them actually get in contact, one obtains
$\sim10^2$ of such 
nanoscale contacts for a visible one. 
Following Refs.~\cite{Zaitsev2008,Tournat2004} we conclude that contribution of such loose but numerous nanocontacts can dominate over the nonlinearity of much stronger (and thus less nonlinear) macrocontacts creating the material skeleton. This explains why nonlinearity can drop drastically after fairly weak shocks 
  that still leave the material skeleton intact, but 
suffice to break the 
nanocontacts. 
 	
{Contour arrows in Fig.\,2(d) schematically} show the physical meaning of the relaxational closing of open contacts and the destructive action of perturbing weak tensile shocks, which do not completely get the system out of the bistability region. We recall that even for large nanocontacts (with $r\agt10^2$\,nm), the bistality zone is of order of atomic size. 
Note that for characteristic attempt times $\tau_0\sim10^{-12}$s and  waiting times below tens of hours, only ``active'' contacts with barriers $ \alt 45k_{B}T$ can participate in thermally induced transitions.   
 Then the narrow ``active'' parts of the energy curves near the bistability-region boundaries can be fairly well approximated by straight segments [thick solid lines in Fig.\,2(d)]. Consequently,  in such narrow regions for almost arbitrary distributions of the asperities' heights, the density of energy states for the active 
nanocontacts can be approximated as constant. 

{\textit{ Kinetic Monte Carlo approach.--} \textcolor{black}{Using 
a kinetic Monte Carlo 
 approach, we simulated  transitions 
 between
 ``open'' or ``closed'' states [see Figs.\,2(b) and 2(d)]  for $3\times10^4$ contacts. 
The probabilities of interstate jumps are given by the aforementioned Arrhenius law.}
If initially all 
nanoscale contacts are broken,   the broken-contact density $N_{b}(E)=1$ and  population of closed 
contacts is zero, $N_{c}(E)=0$. Gradual closing of the broken contacts starts 
  from smallest energy barriers and looks as the 
 motion of the steplike curve $N_{b}(E)$ -- ``closing front'' --  towards the right boundary of the bistability zone with larger barriers. As argued above, the nonlinear-signal amplitude is proportional to the number of closed nano-contacts $\int N_{c}(E)dE$. 
 Curves 1 in  Figs.\,3(b) and 3(d) show the closing front positions after $30$\,s and $1500$\,s of the initial relaxation, respectively. For contacts corresponding to the current position of the relaxation front that 
 moves from the left boundary of the bistability zone to the right \textcolor{black}{[Figs.\,3(c) and 3(d), curves 1]}, the energy wells are 
deep for the open states and shallow for closed ones [see Fig.\,2(d), left side].
 \begin{figure}
\includegraphics[width=8.1 cm]{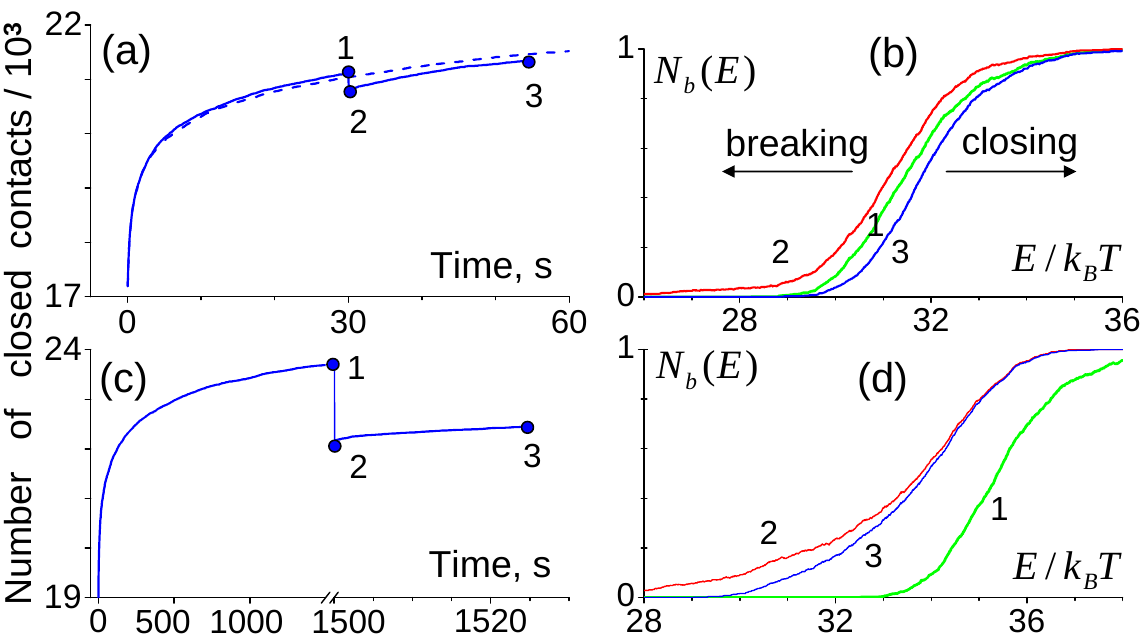}

\caption{(color online). Evolution of the number of closed 
\textcolor{black}{nano-scale contacts (left) and populations $N_{b}(E)$ of broken ones (right). Upper and lower rows are for the  just prepared packing with 
$N_{b}(E)=1$ relaxed during $30$\,s and 1500\,s, respectively. In both cases the same fairly weak perturbing shock that breaks only a portion of earlier closed contacts is applied.
Labels\,1 and 2 are for the system state just before and just after the shock. Labels\,3  are for the system states $24$\,s 
after the shock.}
Compare experimental Fig.\,1(d) with Fig.\,3(a), where the dashed curve shows continuous relaxation. Note the much smaller difference between 
states\,2 and 1 
for the initial relaxation time $30$\,s [(a) and (b)]  than for $1500$\,s [(c) and (d)].}
\label{fig-tip-inter}
\end{figure} 
\textcolor{black}{Sufficiently strong tensile  
perturbations temporarily shift the nanocontacts 
towards the right 
boundary of the energy diagram where, in contrast, energy wells for closed states are shallow [Fig.\,2(d), right side]. 
Consequently,} transitions of previously closed contacts back to open states are 
fostered. 
As a result, at the end of a shock, the preshock 
position of the closing front [Figs.\,3(b) and 3(d), curves 1] 
is shifted back to the left [Figs.\,3(b) and 3(d), curves 2], but 
then the closing front continues its movement to the right. For the same
 shock duration and amplitude, the resulting amount of the broken contacts that increases 
$N_{b}(E)$ essentially depends on the position of the closing front just before the shock \textcolor{black}{(compare the upper and lower rows in Fig.\,3)}.  

\textcolor{black}{Even stronger perturbations can already} shift all bistable contacts to the right beyond the bistability zone 
and break all earlier closed contacts.  
Since the width of the bistability zone is of the order of an atomic size, such strong shocks correspond to strains $\agt10^{-6}-10^{-5}$ for millimetric grains.
For weaker shocks, the system response
can be 
rather multivariant depending on shock amplitude, duration, and previous history.
For example, besides the difference in \textcolor{black}{the amounts of broken contacts}, Fig.\,3(b) shows that, by the same post-shock relaxation time, the closing front (curve 3) gets  already to the right from its initial position [curve 1 in Fig.\,3(b)] 
in contrast to the opposite situation in Fig.\,3(d).           
Next, let us recall that besides the difference in the barrier energies $E_{b,c}$
the widths of bistability zones can strongly differ for different contact sizes. 
Depending on that width, the same perturbation can be ``strong" or ``weak," 
so that perturbation or relaxation regimes for such fractions
of bistable elements 
are quite different. 
       
For the 
discussed features, the ``aging'' of the relaxation response to repeated weak shocks  breaking small contact portions is a natural consequence. Such multiple 
weak
shocks applied to previously well-relaxed material [like in Figs.\,3(c) and 3(d)].
gradually shift the system state towards the one shown in Figs.\,3(a) and 3(b). However, 
that state, for which the relaxed front 3 gets to the right from the initial
position 1, is not reached and the system response saturates 
 when the relaxation between the shocks  becomes able to heal the perturbation
 $\Delta N_{b}$ produced by every previous shock. This saturated value $\Delta N_{b}$ 
 is significantly smaller than the initial reaction 
 of the well-relaxed system to the first perturbing impact. 
\begin{figure}
\includegraphics[width=8.5 cm]{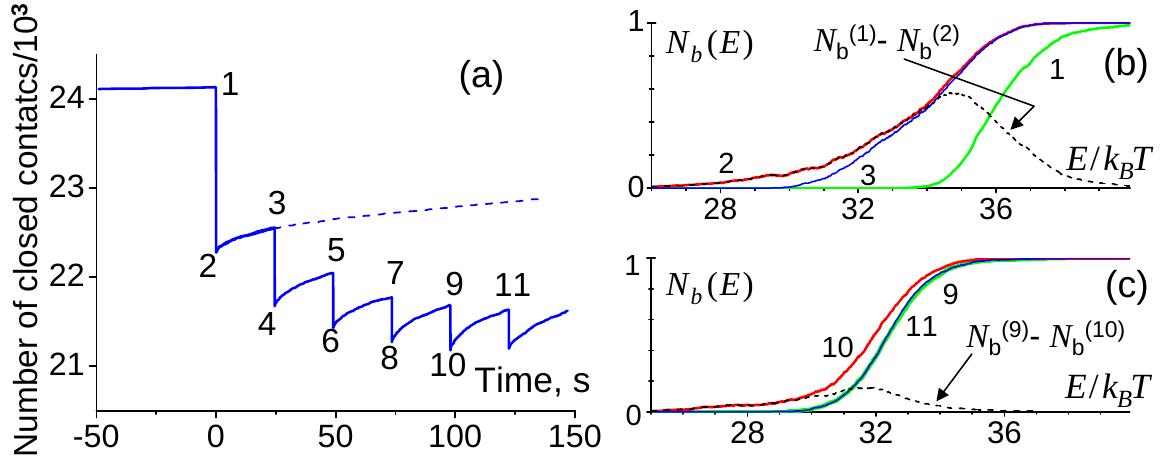}
\caption{(color online). \textcolor{black}{Simulated 
``aging'' of reaction to multiple weak shocks with accumulation of broken contacts for a well relaxed (during $3\times 10^3$s) system.} 
(a) is the time-domain representation similar to the experimental 
 Fig.\,1(c); (b) and (c) are the energy spectra of broken contacts. The \textcolor{black}{mutually-corresponding} moments in plots (a), (b), and (c) \textcolor{black} {are marked by the same numbers. Dashed lines in plots\,(b) and (c) show} 
that initial perturbation $N_{b}^{(1)}(E)-N_{b}^{(2)}(E)$  
strongly exceeds
$N_{b}^{(9)}(E)-N_{b}^{(10)}(E)$ in the ``aged'' state, \textcolor{black}{and the 
``aged'' relaxed 
spectra} $N_{b}^{(9)}(E)$ and $N_{b}^{(11)}(E)$ nearly coincide.  
}
\label{fig-aging}
\end{figure}
The transition to such ``aged'' reaction of the system is shown in Fig.\,4 for the simplest situation of identical 
nanoscale contacts for which the barriers
differ only by initial separation $A$.
This model 
fairly well reproduces  the observed gradual ``aging'' of the system response  
to repeated weak shocks [see Fig.\,1(c)].

{\textit{ Discussion and conclusions.--}} 
Even without 
externally applied shocks, the discussed mechanism of nanocontact destruction or restoration  manifests itself
 in slowly tilted granular packings \cite{Zaitsev2008, JSTAT2010},
where the avalanche precursors act as
internal ``shocks'' followed by relaxations 
[see inset in 
Fig.\,1(b)].
For periodical 
forth-and-back tilting below the critical angle, the demodulated-signal variations  strongly decrease (become ``aged''). However, 
 after $\sim1$\,hour rest, \textcolor{black}{during which the broken 
nanocontacts restore,}  the 
 signal variations also significantly restore \cite{JSTAT2010}. 

\textcolor{black}{
Above we focused on fairly weak perturbations, after which 
the contacts relax independently of each other.} 
If a stronger shock breaks a significant portion of nanocontacts, the surfaces of 
macroscopic intergranular  contacts can experience 
separation 
even greater than 
individual bistability-zone widths. 
Then the interstate jumps of 
nanocontacts 
are not independent, and 
really collective hysteretic or relaxational mechanisms become important
as will be discussed elsewhere. 
But even for weak perturbations,
the revealed mechanism 
demonstrates a rich variety of regimes, including the aging 
 \cite{Amir2011} that has a rather general nature. 
	{The considered bistability of 
	nanoscale contacts is a universal feature of both nonconsolidated (sand-like) and cemented (sandstone-like) granular materials, as well as solids with
	cracks 
	having 
	contacts at their interfaces. This explains universality 
	of slow-dynamics effects observed in those materials \cite{PAJ-Nat2005,Tencate2000,averb-leb2009,Zaitsev2005,Zaitsev2008,JSTAT2010,Guyer2009}.  
Understanding of the gradual accumulation of broken  contacts \textcolor{black}{due to multiple weak perturbations, like in Fig.\,4(a),} opens prospects for physical interpretation of 
	such intriguing phenomena as dynamic earthquake triggering that is phenomenologically discussed in Ref.~\cite{PAJ-Nat2005} and the influence of fairly weak seismo-acoustic stimulation on oil recovery from nearly depleted wells \cite{Beresnev1994}.

Support of ANR Grant 
No. 2010-BLAN-0927-01 and Grant No. 11.G34.31.0066 
of the Russian Federation Government is acknowledged. V.Z. acknowledges the invited-professor grant from the University Rennes-1. We thank J-F.\,Bardeau for the help with AFM imaging.



\end{document}